\documentclass{IET}
\usepackage{algorithm}
\usepackage{algorithmic}
\usepackage{graphicx}
\usepackage{amsmath}
\usepackage{multirow}
\begin{document}
\title{Low-complexity feedback-channel-free distributed video coding using Local Rank Transform}

\author{P Raj Bhagath}
\affil[1,*]{Department of Electronics and Electrical Communication Engineering, Indian Institute of Technology, Kharagpur, INDIA (e-mail: rajbhagath@gmail.com)}

\author{Kallol Mallick}
\affil{Azure Software Systems Ltd., Kolkata, INDIA (e-mail: kallol.mallick@gmail.com)}

\author{Jayanta Mukherjee}
\affil{Department of Computer Science and Engineering, Indian Institute of Technology, Kharagpur, INDIA (e-mail: jay@cse.iitkgp.ernet.in)}
\author{Sudipta Mukopadhayay}
\affil{Department of Electronics and Electrical Communication Engineering, Indian Institute of Technology, Kharagpur, INDIA (e-mail: smukho@ece.iitkgp.ernet.in)}
\affil[*]{rajbhagath@gmail.com}

\abstract{In this paper, we propose a new feedback-channel-free Distributed Video Coding (DVC) algorithm using Local Rank Transform (LRT). The encoder computes LRT by considering selected neighborhood pixels of Wyner-Ziv frame. The ranks from the modified LRT are merged, and their positions are entropy coded and sent to the decoder. In addition, means of each block of  Wyner-Ziv frame are also transmitted to assist motion estimation. Using these measurements, the decoder generates side information (SI) by implementing motion estimation and compensation in LRT domain. An iterative algorithm is executed on SI using LRT to reconstruct the Wyner-Ziv frame. Experimental results show that the coding efficiency of our codec is close to the efficiency of pixel domain distributed video coders  based on Low-Density Parity Check and Accumulate (LDPCA) or turbo codes, with less encoder complexity.}

\maketitle

\section{Introduction}
The emerging applications like mobile camera phone, video surveillance, multimedia sensor networks, etc., demand a low cost encoder with high coding efficiency. This is due to the fact that the encoder has less memory and less computational power. Some applications like video storage based applications and real-time video streaming over Internet does not have feedback channel. DVC is a coding paradigm, which provides the potential of shifting encoder complexity to decoder while achieving coding efficiency near to conventional video coding techniques. DVC is based on two well known information theoretic results namely, the Slepian-Wolf (SW) theorem and Wyner-Ziv (WZ) theorem [1,2].

In most of the existing DVC schemes, the video frames are divided into key and Wyner-Ziv (WZ) frames.
The key frames are intra coded using conventional video coding methods. The WZ frames, coded using LDPCA or turbo codes and the parity bits are stored in a buffer. The parity bits are sent to the decoder on request. At the decoder, the Side Information (SI) is generated using Motion Compensation Interpolation (MCI) technique [3]. The WZ frames are then decoded by joint processing of the SI and parity bits from the encoder.
The MCI technique generates SI without using any information from the current frame at the decoder.
The SI generated by MCI technique is of poor quality, and to improve it, a successive refinement of WZ frames is proposed in [3]. The practical implementation of DVC codecs, in pixel domain is proposed in [4,5]. DVC in transform domain using LDPCA codes is proposed in [6]. In PRISM [7], blocks are classified as skip, intra or WZ, and coded accordingly.

Distributed video coding using LRT is first proposed in [8]. It reports less computational complexity in LRT encoder over
conventional DVC methods like LDPCA and turbo codes. It includes a novel approach for generating SI using ME and MC in LRT domain, and also by performing the image reconstruction using a regularization technique, named DLRTexReg [8]. However, the scheme reports relatively poor rate distortion performance compared to conventional LDPCA based codecs.

In our approach, we compute LRT using a subset of neighborhood pixels to reduce the bit rate and encoder computational complexity with a negligible loss in PSNR of reconstructed image. In addition, mean pixel intensity of each block of WZ frame is sent to the decoder to assist motion estimation, which improves the SI quality. To further reduce the bit rate, sampled LRT is used. We modify the decoding scheme reported in [8] to process the sampled LRT of an image. In the modified algorithm,  the unknown rank pixel intensity values are recovered using SI and known rank pixel intensity values. The proposed codec does not require any feedback channel, as we are coding all rank values independent on the response from the decoder. In the next section, we briefly discuss the base algorithm as reported in [8]. Subsequently we present the modifications proposed in this work for improving its performance.

\section{The Base Algorithm}
\begin{figure*}[hpt]
\begin{center}
\centering{\includegraphics[width=18cm,height=5.76cm]{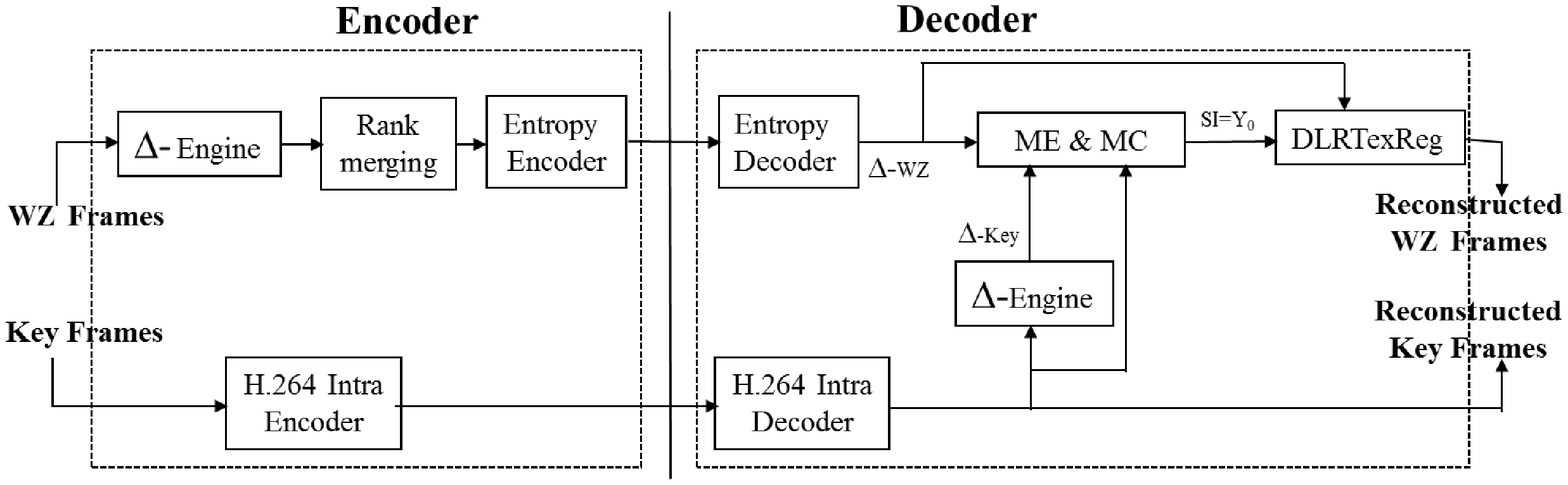}}
\caption{Block diagram of DVC codec using LRT.}
\label{fig:scheme1}
\end{center}
\end{figure*}
The block diagram of the DVC codec using LRT, which is proposed in [8] is shown in Fig. 1. At the encoder, each input frame is treated either as a key frame or a WZ frame. Key frames are H.264 intra encoded and for WZ frames, LRT is computed and the obtanied rank values are entropy coded. The basic definitions for LRT are provided below.

In [9], Local Rank Transform (LRT) of a set $S$ is defined as
\begin{equation}
LRT(S)=\{r(x;\aleph(x))|x\epsilon S\}
\end{equation}
where,

$r(x;\aleph(x))$-rank of element $x$ with respect to $\aleph(x)$ [8], and
$\aleph(x)$-neighborhood of $x$ and is subset of $S$.
The rank of $x$ with respect to set $S$ is defined as the number of elements less than $x$ in $S$.

The $\delta$-rank of $x$ with respect to S is defined as the number of elements less than $x$ by at least $\delta$ amount, and is denoted as $r_{\delta}(x;S)$.

$\delta$-Local Rank Transform Extended Neighborhood of an image $I$ is defined in [8], as
\begin{equation}
LRT_{\delta}^{m,n}(I)=\Delta(I)=\{r_{\delta}(x;\aleph^{m,n}(x))|x\epsilon I\}
\end{equation}
 where,

$\aleph^{m,n}(x)$- extended neighborhood of $x$.

Usually, we use $m=n=N$ for a block. So, the above mentioned $\aleph^{m,n}(x)$ is denoted as $\aleph^{N}(x)$.
For WZ frames, $\delta$-LRT is computed over a neighborhood size $N\times N$ and the rank values are sampled, merged, compacted and entropy encoded.
All pixel ranks are considered in image reconstruction for high bit rates, where as only selected pixel ranks are considered for low bit rates for different values of $N$ $(N=1,2,...)$. We denote the transform containing all pixel ranks as \textit{full LRT} and selected pixel ranks as \textit{sampled LRT} in subsequent sections.

The decoding algorithm starts with entropy decoder to produce the LRT image back. In case of sampled LRT, the decoder interpolates the missing rank values from neighboring rank values. Then ME and MC are performed in LRT domain to generate the SI, which is used as initial estimate for WZ frame reconstruction with the DLRTexReg algorithm [8]. In this algorithm, for each pixel $x$ of SI, a cost function ($E_{data}(x)+\lambda E_{prior}(x)$) is computed, when $x$ is updated (increased/decreased by a constant, or unchanged). $E_{data}(x)$ is given by,
\begin{equation}
E_{data}(x)= |lr_{\delta}(x)-lr_{\delta}(x^{'})|
\end{equation}
Where $lr_{\delta}(x)$ is the $\delta$-local rank of $x$ and $x^{'}$ is the new value of $x$.
$E_{prior}$ is a smoothness term, which measures the difference between the statistical mean of co-located pixels in the neighborhood $(\mu^{c})$.
\begin{equation}
E_{prior}(x)= |\mu(x,N) - \mu^{c}(x,N,\hat{I})| \hspace{1cm}\forall x\epsilon I_{0}
\end{equation}

where $\mu(x,N)$ is mean of the intensity values of the pixels in neighborhood (N) of $x\epsilon I_{0}$, and $\mu^{c}(x,N,\hat{I})$ is the mean calculated over neighborhood of the corresponding co-located pixel in image $\hat{I}$. $I_{0}$ and $\hat{I}$ represent SI and reconstructed image, respectively. The option  which produces lowest value of the cost function is used to update $x$.

DLRTex is the basic variant of DLRTexReg algorithm, and is used in our proposed scheme. DLRTex algorithm takes SI and original rank values of WZ frame as inputs. Firsty, it computes $\delta$-LRT of SI with neighborhood size $N$, and compares the rank value of each pixel with the original rank value. If the calculated rank value is less than the original rank value, then the pixel intensity value is increased by a constant $step$, whereas if it is greater, the pixel intensity value is decreased by $step$. If both the rank values are same then the pixel intensity value remains unchanged. The algorithm iteratively updates SI by comparing rank values. In each iteration, Peak Signal to Noise Ratio (PSNR) value between original rank image and updated SI rank image is computed. If this value is less than the PSNR value in the previous iteration then the algorithm stops. The pseudo code for DLRTex algorithm is given in Algorithm 1.

\begin{algorithm}{DLRTex}
\textit{Algorithm 1}: \textbf{DLRTex} \newline
\newline
\textbf{Input}:  $\Delta(I)$: $\delta$-LRT with N neighborhood and $\delta$. \newline
     \hspace{3cm}  $I_{0}$:Initial estimate image. \newline
\textbf{Output}: $\hat{I}$: Reconstructed image. \newline
\textbf{Parameters}:  N: Neighborhood size for LRT: n=m=N. \newline
            $\delta$: $\delta$ value to use in  $\delta -$ LRT. \newline
            \textit{step}: incremental (or decremental) update factor. \newline

Let us define \newline
$I^{i}_{Est}$: reconstructed image at $i$-th iteration.\newline
$Y^{i}(x)$: intensity at pixel $x$ in $i$-th iteration
$\forall x \epsilon I^{i}_{Est}$ and \newline

\textbf{begin} \newline
1. Set $R^{Ref}=\Delta(I)$, such that $R^{Ref}(x)=r_{\delta}(x),  \forall x \epsilon I.$ \newline
2. $I^{0}_{Est}=I_{0}$ and $R^{0}=\Delta(I^{0}_{Est})$. \newline
3. In the $i$-th iteration \newline
\textit{for each pixel} $x$ in  $I^{i-1}_{Est}$ \newline
Assume, $x^{'}$ is the co-located (with $x$) element in $R^{Ref}$

\begin{algorithmic}
        \IF{$R^{Ref}(x^{'}) > R^{i}(x)$}
          \STATE $Y^{i}(x)=Y^{i-1}(x)+step$.
        \ELSIF{$R^{Ref}(x^{'}) < R^{i}(x)$}
         \STATE     $Y^{i}(x)=Y^{i-1}(x)-step$.
        \ELSE
          \STATE  $Y^{i}(x)=Y^{i-1}(x)$.
          \ENDIF \newline
    \end{algorithmic}
\textit{end for loop} \newline
Calculate $R^{i}=\Delta(I^{i}_{Est})$. \newline
Calculate $PSNR(R^{ref},R^{i})$. \newline

4. If $PSNR(R^{ref},R^{i})$ is less than $PSNR(R^{ref},R^{i-1})$, set  $\hat{I}$=$I^{i-1}_{Est}$ and stop iteration.
\end{algorithm}

\section{Proposed Modifications}
In this section, we describe the proposed modifications to the base algorithm in detail.

\subsection{Encoder}
In our scheme, we use a modified LRT to decrease the bit rate. The transformed rank values are merged to decrease the bit rate further, and their positions are entropy coded. The following subsections describes the modified LRT and rank merging in detail.

\subsubsection{$\Delta_{even}$ or $\Delta_{odd}$}
\begin{figure}[h]
\centering{\includegraphics[width=5.7cm,height=2.85cm]{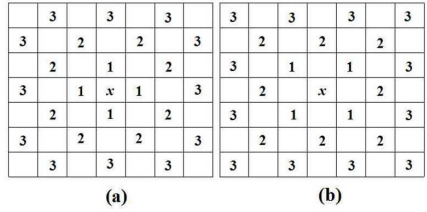}}
\caption{(a). Odd and (b). Even neighbor pixels of pixel $x$.}
\label{fig:LRTS}
\end{figure}

The neighborhood pixels of a pixel $x$ ($\aleph^{N}(x)$) are divided into two mutually exclusive and exhaustive subsets namely, $\aleph^{N}_{odd}(x)$ and $\aleph^{N}_{even}(x)$ as shown in Fig. 2.
If the city block distance from $x$ to a pixel is odd, then the pixel is considered as an odd neighbor of $x$.
Similarly, if the city block distance from $x$ to a pixel is even, then the pixel is an even neighbor of $x$. For neighborhood size $N=1$, pixels labeled as `1' in Fig. 2, and for neighborhood size $N=2$,
pixels labeled as `1' and `2' are considered as neighborhood pixels (odd and even). Neighborhood pixels are chosen in the same manner for the higher values of $N$.

Now, we define  the $\delta$-LRT extended odd neighborhood ($\Delta_{odd}$) of image $I$ as
\begin{equation}
\Delta_{odd}(I)=\{r_{\delta}(x;\aleph^{N}_{odd}(x))|x\epsilon I\}
\end{equation}
and the $\delta$-LRT extended even neighborhood ($\Delta_{even}$) of image $I$ as
\begin{equation}
\Delta_{even}(I)=\{r_{\delta}(x;\aleph^{N}_{even}(x))|x\epsilon I\}
\end{equation}

In case of $\Delta_{odd}(\Delta_{even}$) the maximum rank value $|S|$ is less compared to $\Delta$. For different values of $N$
the maximum rank values are given in Table \ref{table:t1}.
\begin{table}
 \caption {Maximum rank value $|S|$ comparison}
 \label{table:t1}
 \begin{tabular}{l|p{1cm}|p{1cm}|p{3cm}|}
 \cline{2-3}
 & \multicolumn{2}{ |c| }{Maximum Rank value $|S|$} & \\ \cline{1-4}
 \multicolumn{1}{ |c| }{N} & $\Delta$ & $\Delta_{odd} (\Delta_{even})$ & No. of Distinct Rank values after merging of $\Delta_{odd} (\Delta_{even})$ \\  \hline
 \multicolumn{1}{ |c| } {1} & 8 & 4 & 3 \\ \hline
 \multicolumn{1}{ |c| } {2} & 24 & 12 & 9\\ \hline
\multicolumn{1}{ |c| } {3} & 48 & 24 & 18\\ \hline
\multicolumn{1}{ |c| } {4} & 80 & 40 & 30\\ \hline
 \end{tabular}
 \end{table}
The WZ frames are either $\Delta_{odd}$ or $\Delta_{even}$ transformed with parameters N (neighborhood size) and $\delta$.

\subsubsection{Merging Ranks and Entropy encoding}
The ranks are merged to decrease the bit rate without losing quality of the reconstructed image. Histogram of $\Delta_{odd}$ rank image for different values of $N$ is shown in Fig. 3. From Fig. 3 we can observe that, the lower half ranks have less probability of occurrence compared to higher half. So, consecutive ranks of lower half are merged, and replaced with the higher rank value between them, as shown in Fig. 4. The number of edge pixels is less than the number of smooth region pixels of an image. Smooth regions of an image have the highest rank for negative values of $\delta$ [9]. So, the probability of pixels with higher rank is high compared to other rank pixels (see Fig. 3). For different sequences, the average number of bits reduced by merging ranks is shown in Table \ref{table:t2}.
\begin{figure}[h]
\centering{\includegraphics[width=8cm,height=4.2cm]{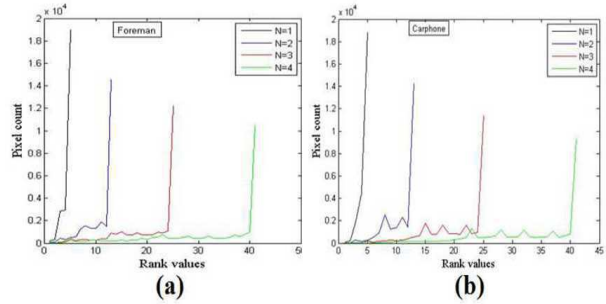}}
\caption{Histogram of $\Delta_{odd}$ of WZ frame before merging ranks for a). Foreman, b). Carphone sequence.}
\label{fig:hist}
\end{figure}

\begin{figure}[h]
\centering{\includegraphics[width=8.5cm,height=2.08cm]{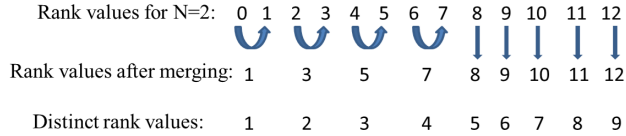}}
\caption{Rank merging for neighborhood size 2.}
\label{fig:merging}
\end{figure}
\begin{table}
 \caption {Comparison of PSNR and number of bits for different sequences without and with rank merging}
 \label{table:t2}
 \begin{tabular}{l|p{1cm}|p{1cm}|p{1cm}|p{0.8cm}|p{1.2cm}|}
 \cline{2-5}
 & \multicolumn{2}{ |c }{Without Merging} & \multicolumn{2}{ | c |}{With Merging}\\ \cline{2-6}
 \multicolumn{1}{ c| }{} & PSNR (dB) & No.of bits & PSNR (dB) & No.of bits & \% bits reduced \\  \hline
 \multicolumn{1}{ |c| } {Foreman} & 37.61 & 50,375 & 37.35 & 40,817 & 18.97 \\ \hline
 \multicolumn{1}{ |c| }  {Carphone} & 38.36 & 45,416 & 38.08 & 36,648 & 19.31\\ \hline
\multicolumn{1}{ |c| }  {coastguard} & 36.60 & 73,812 & 36.40 & 60,428 & 18.13\\ \hline
 \end{tabular}
 \end{table}
$\Delta_{odd}$ or $\Delta_{even}$ image has maximum rank value $|S|<<255$, which depends on the value of N.
So instead of coding rank values directly, their positions are entropy coded which enables flexibility in rank merging. After merging the ranks, their positions are represented using binary values as explained in Fig. 5. Staring with highest rank, all the rank positions are represented with `1' and their absence is represented with `0'. If a rank is coded once then its position is not considered for the next rank coding. These binary values are entropy coded using MQ coder [10].

Context used here is calculated as number of pixel positions already coded in the 8 neighborhood of the pixel under consideration. Whenever a rank position is coded, its status is updated in temporary variable. Initially status of all pixel positions are assigned to zero. Once a particular rank position is coded then its status is updated to 1. This context information along with binary value of pixel is sent to MQ-coder. Let $a,b,c,d,e,f,g,$ and $h$ denote the status of 8 neighbors of pixel $x$ then context of $x$ is given by $CX(x)=a+b+c+d+e+f+g+h$. Number of contexts used in this model are 9.

Mean intensity values of $16\times 16$ block are also sent to the decoder to assist Motion Estimation. The overhead of sending mean values of each block is marginal (792 bits per frame for qcif sequence) as we are considering large block size, and it shows significant improvement in PSNR of reconstructed image (Table \ref{table:t3}).

\begin{figure}[h]
\centering{\includegraphics[width=8.5cm,height=7cm]{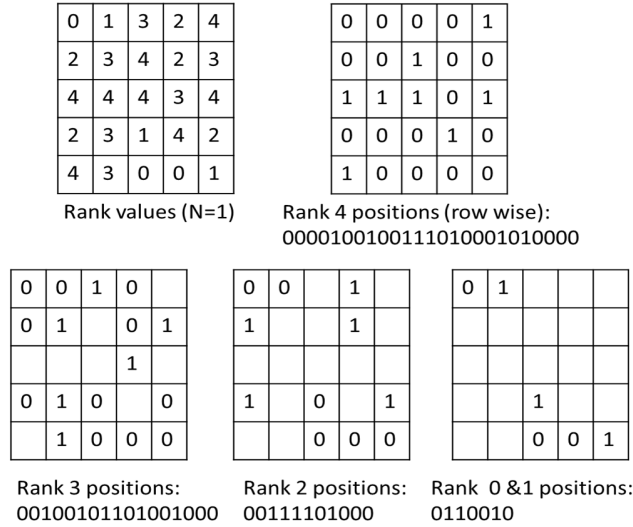}}
\caption{Example for position coding of merged ranks.}
\label{fig:poscod2}
\end{figure}

\subsection{Decoder}

First stage of decoding algorithm is entropy decoder to get the $\Delta_{odd}$ or $\Delta_{even}$ of WZ frame back. SI is generated by ME and MC, and it is used as initial estimate ($I_{0}$) for image reconstruction using DLRTex. For more accurate motion estimation, we use mean value of each block additionally as discussed in the following subsection.

\subsubsection{Mean assisted ME and MC}
For each $16\times 16$ block of WZ frame, ME and MC are performed on key frames to get SI.
ME and MC in LRT domain are proposed in [8], where sum of absolute difference (SAD) of ranks is used as a measure in ME.
Blocks with the least SAD value of ranks are considered for ME. LRT image contains the edge information. But in an image, some portions may have same edge information with different contrast which may mislead ME. For example, in set $S_{1}$=\textit{\{20, 25, 18, 39, 9\}}, the rank of the third element (18) is 1. In set $S_{2}=2S_{1}+110$=\textit{\{150, 160, 146, 188, 128\}}, the value of the third element changes to 128, but its rank remains as 1. So, mean intensity value of each block is used to assist ME.
For every $16\times16$ block of WZ frame, ME is done in LRT domain as follows:

Two motion vectors $MV_{1}$ and $MV_{2}$ are computed.
\begin{enumerate}
\item $MV_{1}$- motion vector with least SAD of ranks $LSAD_{1}$.
\item $MV_{2}$- motion vector with mean intensity difference between blocks less than a threshold ($T_{1}$) and least SAD of ranks $LSAD_{2}$.
\end{enumerate}

The absolute difference between $LSAD_{1}$  and $LSAD_{2}$ is the key to choose relevant motion vector. For background and low motion blocks, mean difference condition always satisfies. That means, within the search region there is at least one block which satisfies the mean difference condition and with less error between ranks. The difference between $LSAD_{1}$ and $LSAD_{2}$ is zero or low for such blocks. For high motion or newly exposed region blocks, the mean difference condition may not satisfy as the matched block may not be available within the search region. In such cases the difference between $LSAD_{1}$  and $LSAD_{2}$ is large, and it would be better to choose motion vector with least SAD value, which at least contain edge information (i.e. $MV_{1}$).

\begin{equation}
MV=\begin{cases} MV_{1} &\text{if} \hspace{1em} |LSAD_{2}-LSAD_{1}|>T_{2},\\
MV_{2} &\text{otherwise}.
\end{cases}
\end{equation}

where, $T_{2}=0.05*|S|*(16*16)$ and $T_{1}=5$, chosen experimentally. Motion compensation is done using $MV$ to generate SI. In case of \textit{sampled LRT}, the constant $T_{2}$ is reduced by half as the number of pixels also decreases by half.

Blocks encircled in red in Figs. 6(b) and 7(b) have same ranks as the original image blocks, but their mean values are different. Even after reconstruction using DLRTex the errors remain.

Table \ref{table:t3} presents the PSNR values of SI and reconstructed images without and with using mean assistance in motion estimation. High motion frames are taken for comparison. From Table \ref{table:t3}, we can clearly observe that the mean assisted motion estimation yields reconstructed images with better PSNR.
\begin{table}
 \caption {PSNR comparison of SI and reconstructed images without and with mean assistance in ME for different values of N.}
 \label{table:t3}
 \begin{tabular}{|l|c|p{1cm}|p{1cm}|c|}
 \hline
 Sequence(frame number) & N & PSNR (dB) & Without mean & With mean \\ \hline

 \multirow{6}{*}{Foreman(94)} & \multirow{2}{*}{1} & SI & 23.75 & 28.01 \\ \cline{3-5}
 &  & Rec &  24.68 & 29.36 \\ \cline{2-5}
 & \multirow{2}{*}{2} & SI & 25.41 & 28.54 \\ \cline{3-5}
 &  & Rec &  26.83 & 30.93 \\ \cline{2-5}
 & \multirow{2}{*}{3} & SI & 26.95 & 28.79 \\ \cline{3-5}
 &  & Rec &  29.16 & 31.96 \\ \hline
 \multirow{6}{*}{Carphone(47)} & \multirow{2}{*}{1} & SI & 32.24 & 33.16 \\ \cline{3-5}
 &  & Rec &  32.67 & 33.71 \\ \cline{2-5}
 & \multirow{2}{*}{2} & SI & 34.37 & 35.60 \\ \cline{3-5}
 &  & Rec &  35.21 & 36.80 \\ \cline{2-5}
 & \multirow{2}{*}{3} & SI & 34.58 & 35.50 \\ \cline{3-5}
 &  & Rec &  35.74 & 37.07 \\ \hline
 \multirow{6}{*}{Coastguard(74)} & \multirow{2}{*}{1} & SI & 27.08 & 27.04 \\ \cline{3-5}
 &  & Rec &  28.11 & 28.20 \\ \cline{2-5}
 & \multirow{2}{*}{2} & SI & 27.17 & 27.28 \\ \cline{3-5}
 &  & Rec &  29.08 & 29.35 \\ \cline{2-5}
 & \multirow{2}{*}{3} & SI & 27.19 & 26.95 \\ \cline{3-5}
 &  & Rec &  29.45 & 29.7 \\ \hline

 \end{tabular}
 \end{table}

\begin{figure}[h]
\centering{\includegraphics[width=8cm,height=3.24cm]{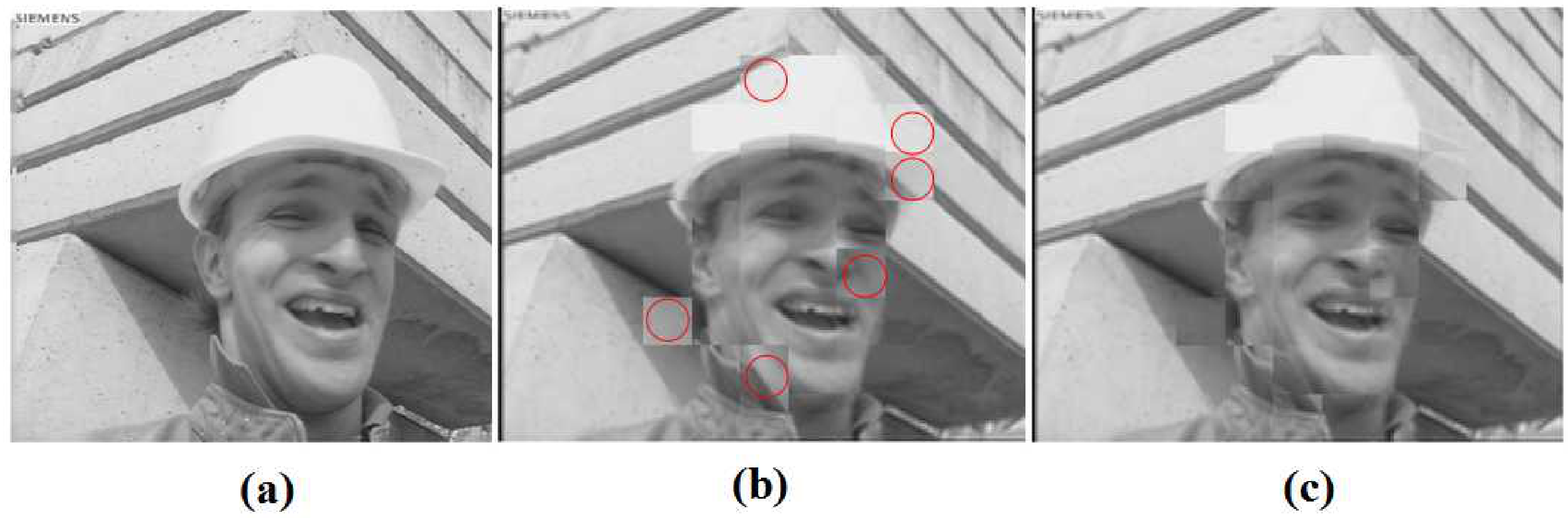}}
\caption{Comparison of (a) original image with (b) generated SI with out mean assistance and (c) with mean assistance for foreman (QCIF) sequence 15th frame.}
\label{fig:LRTS}
\end{figure}

\begin{figure}[h]
\centering{\includegraphics[width=8cm,height=2.46cm]{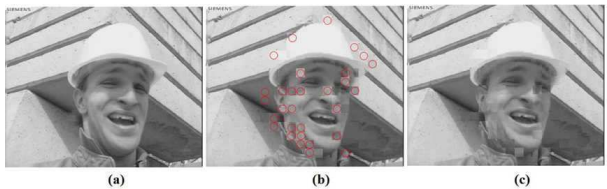}}
\caption{Comparison of (a) original image with (b) generated SI with out mean assistance and (c) with mean assistance for foreman (CIF) sequence 15th frame.}
\label{fig:LRTS}
\end{figure}
Finally, DLRTex is carried out with $\Delta_{odd}$ of WZ frame and SI as inputs. The parameters involved to reconstruct the WZ frames are N, $\delta$ and \textit{step} (incremental/decremental factor) (refer to Section 3).

\subsection{Reconstruction algorithm for Sampled LRT}
\textit{Sampled LRT} contains the information of half of the pixels rank values of WZ frame, and other rank values are unknown.
The reconstruction algorithm is different for pixels with known and unknown ranks. The steps involved are:

a). \textit{Motion estimation and compensation using sampled LRT}:

For the available neighborhood key frames, $\Delta_{even}$ is computed. Mean assisted ME and MC are performed using sampled
LRT to get SI. The generated SI is used as initial estimate for reconstructing the pixels of known ranks.

b). \textit{Reconstruction of pixels with known ranks}:

The pixels with known ranks are reconstructed using DLRTex algorithm with calculated SI and $\Delta_{even}$ as inputs.

c). \textit{Reconstruction of pixels with unknown ranks}:
\begin{figure}[h]
\centering{\includegraphics[width=8cm,height=11.22cm]{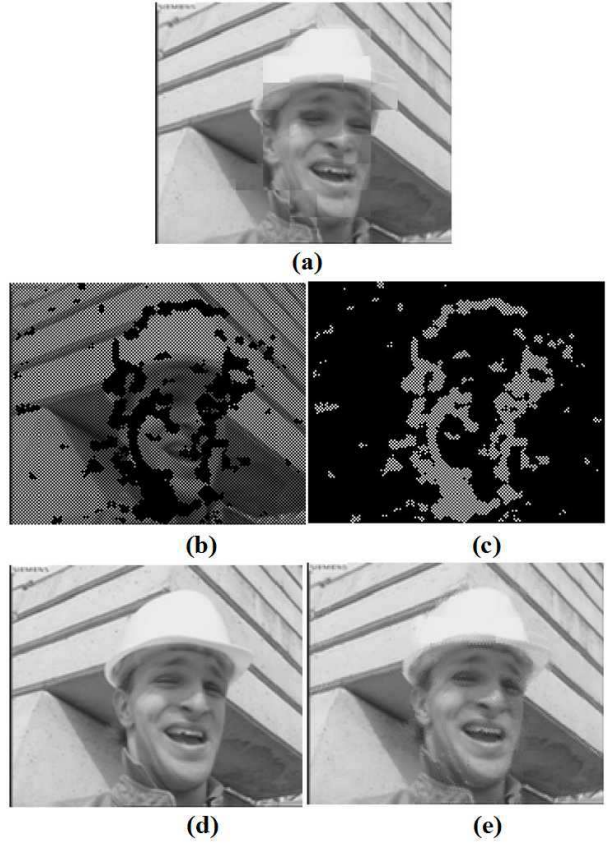}}
\caption{Images at different levels of reconstruction using sampled LRT. a). SI , b). background or low motion pixels copied from SI, c). high motion pixels, d). original image and e). reconstructed image of foreman sequence 15th frame.}
\label{fig:LRTS}
\end{figure}
The pixels with unknown ranks are divided into two groups: background or low motion pixels and high motion pixels.
The pixel can be considered as background or low motion pixel, when its neighborhood pixel ranks match
with corresponding ranks of motion compensated pixels. SAD value is used as a measure of matching.
If SAD value is less than a threshold ($T_{3}$) then the unknown pixel is copied from the co-located  pixel at SI,
else (SAD value $\geq T_{3}$) it is treated as high motion pixel. These pixels are reconstructed by taking
average of their neighborhood (known rank) pixel intensity values.

Fig. 8(a) shows the SI image generated using \textit{sampled LRT} for foreman sequence WZ frame. The PSNR of the SI (29.5 dB) is
slightly less than that of \textit{full LRT} (30.26 dB).
Firstly, background or low motion and high motion pixels of unknown rank pixels are identified.
As explained above, these pixels of unknown rank pixels are copied from SI image as shown in Fig. 8(b).
Using DLRTex the known rank pixel intensity values are reconstructed. The identified high motion pixels are shown in Fig. 8(c), and
their intensity values are calculated by taking average of neighborhood known rank pixel intensity values.
Fig. 8(e) shows the final reconstructed image using \textit{sampled LRT} with N=3.

\subsection{Post processing of decoded frames}

At the decoder side, we have the original image mean intensity values of each block. These mean values are used in ME, and can also be used for post processing the decoded image to improve quality. After decoding the complete image using DLRTex, for each block, mean values are calculated. Let $M_{d}$ denote the mean value of block $B$ in the decoded image, and $M_{o}$ denote its original mean value sent from the encoder. Then each pixel intensity $Y(i,j)$ in block $B$ is updated as follows.

\begin{equation}
\overline{Y}(i,j) = Y(i,j)-M_{d}+M_{o},\hspace{1em} \forall (i,j)\epsilon B
\end{equation}
\begin{equation}
Y_{p}(i,j)=\begin{cases}
0 & if\hspace{1em}  \overline{Y}(i,j)\leq 0,\\
255 & if  \hspace{1em} \overline{Y}(i,j)\geq 255,\\
\overline{Y}(i,j) & otherwise.
\end{cases}
\end{equation}
After updating every pixel in each block, DLRTex algorithm is again performed to remove the blocking artifacts. This post processing improves the PSNR by 0.5dB-2dB depending on block size. If the block size is low then it improves PSNR for all data points with different neighborhood sizes, but bit rate increases as the number of bits to sent the mean values also increases. If the block size is high then only higher data points gets the benefit.

\section{Encoder Complexity}
\subsubsection{Complexity of LRT}
To calculate rank of a pixel, the number of comparisons and incremental operations required is $2N(N+1)$, where $N$ is the neighborhood size. An extra addition operation per pixel is required to calculate the mean value of a block. As the mean value is calculated for a block of 256 pixels, which is a power of 2, the division operation can be performed by simple shifting operation. Hence, for an image with $P$ pixels, the total number of operations required are
\begin{equation}
\Pi_{LRT}(N)=2PN(N+1)(C+I)+PA
\end{equation}
Where $C$, $I$ and $A$ denote comparison, increment and addition operations respectively.
In the case, where neighborhoods overlap, half the comparisons can be avoided, by storing the comparison results in a bit array and re-using them. Hence the total number of operations is updated as
\begin{equation}
\Pi_{LRT}(N)=PN(N+1)(C+2I)+PA
\end{equation}
\subsubsection{Complexity of Context modelling}
In our proposed scheme rank positions are encoded instead of rank values directly. Let $P_{|S|}, P_{|S|-1}, ... P_{0}$ denote the occurrences of rank values from higher rank to the lower.
Rank values are coded from highest to lower. So the total number position symbols to be coded depends up on occurrences of $P_{i}$.
The number of pixels positions to be coded are $P$ for coding highest rank value and $P-P_{s}$ for the next highest rank value  and so on. So the total number of symbols to be coded are

\begin{equation}
\beta =P+P-P_{|S|}+P-(P_{|S|}+P_{|S|-1})+.......
\end{equation}

To calculate the context of a pixel, the number of increment operations required are $4$ for highest rank value and $8$ for other rank values. This increment operation is controlled by 8 neighbours. One comparison operation is required per pixel to search for a particular rank value to be coded. The number of operations required to find the context of highest rank pixels are $P(C+4I)$, and it is $(\beta-P)\times(C+8I)$ for other ranks. So the total number of operations required to find contexts of all rank value positions are

\begin{equation}
\Pi_{CX}=P(C+4I)+(\beta-P)\times(C+8I)
\end{equation}

\subsubsection{Complexity of MQ-encoder}
MQ coder is used to code rank positions. The MQ-Coder utilizes a probability model for its encoding process. This model is implemented as a Finite State-Machine (FSM) of 47 states. Two memory read operations are performed per symbol to read probability value of LPS ($Q_{e}$) and MPS sense [10] and one comparison operation to find expected symbol. One addition and one move operations are required for updating A and C registers. These registers are normalised, if A falls below a certain threshold. Normalisation requires at max two shifting operations. Register CT gets decremented by one whenever shifting occurs in normalisation. Byte out procedure is executed whenever CT reaches zero. Byte out procedure does one memory write operation, and resets counter CT.
The maximum number of operations performed per symbol is $3M+C+A+2S_{H}+D+M_{V}$. $S_{H}$ and $M_{V}$ indicate shift and move operations. So the total numbers of operations performed by MQ-encoder are
\begin{equation}
\Pi_{MQ}=\beta \times(3M+C+A+2S_{H}+D+M_{V}).
\end{equation}

Total cost of the encoder is
\begin{equation}
\Pi_{total}=\Pi_{LRT}+\Pi_{CX}+\Pi_{MQ}
\end{equation}

In case of LDPC, to generate n-th code, the computational complexity [8] for a frame of QCIF size is
\begin{multline}
\Pi_{LDPC}(n)=8P\times (4A+7M+D+\frac{n+1}{132} (DIV+MULT+2A+D))
\end{multline}
Where, $M$, $X$, $DIV$, and $MULT$ denote memory copy, modulo-2 operation, division and multiplication, respectively.
In an embedded domain (Intel Atom$^TM$ processor), the average number of clock cycles required [12] per instruction would be $C$ = $A$ = 1, $DIV$ = $MULT$ = 6, $I$ = $S_{H}$ = $D$ = $X$ = $M_{V}$ = 0.5, and $M$ = 3. LRT computation involves simple comparison, increment and addition operations, which are responsible for its less complexity when compared to LDPC.

Power required to transmit WZ frames in watts is given by
\begin{equation}
P_{WZ}=k(f_{WZ}\Pi + \alpha R_{WZ})
\end{equation}

Where $k=CV^{2}\mu_{s}$ is processor technology dependent constant (chosen value of $k$ is 1 so that power values we get are some scaled values. Typically, for 0.18nm technology C=0.015pf and V=1.8v. $\mu_{s}$ is average switching per clock cycle and depends on the data switching. So the power we have compared is in the order of milli watts), $f_{WZ}$ is WZ frame rate (15 frames per second), $\Pi$ is complexity of the encoder in terms of number clock cycles, $\alpha$ is the power ratio between transmitting power and processing power ($\alpha= 50$) and $R_{WZ}$ is the bit rate of WZ frames. Per frame power comparison of LRT based encoder and LDPC based encoder is given in Tables. \ref{table:t4} and \ref{table:t5}. From tables it is observed that LRT based WZ frame encoding has lower power consumption compared to LDPC based DVC.

\begin{table}
 \centering
 \caption {Relationship between average PSNR, rate and power for LRT encoder for foreman sequence}
 \label{table:t4}
 \begin{tabular}{c p{2cm} p{2cm} p{2cm} p{2cm}}
 \hline
 N & Sampling ratio & PSNR(dB) & Rate(kbps) & power(scaled) (unit)\\
 \hline
 1 & 0.5 & 30.57 & 202 & 172\\
 2 & 0.5 & 32.35 & 265 & 244\\
 3 & 0.5 & 33.56 & 344 & 381\\
 1 & 1 & 32.19 & 280 & 276\\
 2 & 1 & 34.02 & 403 & 418\\
 3 & 1 & 36.60 & 630 & 729\\
 \hline
 \end{tabular}
 \end{table}

 \begin{table}
 \centering
 \caption {Relationship between average PSNR, rate and power for LDPC encoder for foreman sequence}
 \label{table:t5}
 \begin{tabular}{c p{2cm} p{2cm} p{2cm}}
 \hline
 n & PSNR(dB) & Rate(kbps) & power(scaled) (unit)\\
 \hline
 8 & 28.30 & 180 & 923\\
 11 & 30.70 & 247 & 967\\
 15 & 33.00 & 337 & 1027\\
 23 & 35.50 & 517 & 1146\\
 \hline
 \end{tabular}
 \end{table}
 \begin{figure*}[hpt]
 \begin{center}
 \centering{\includegraphics[width=13.09cm,height=6.6cm]{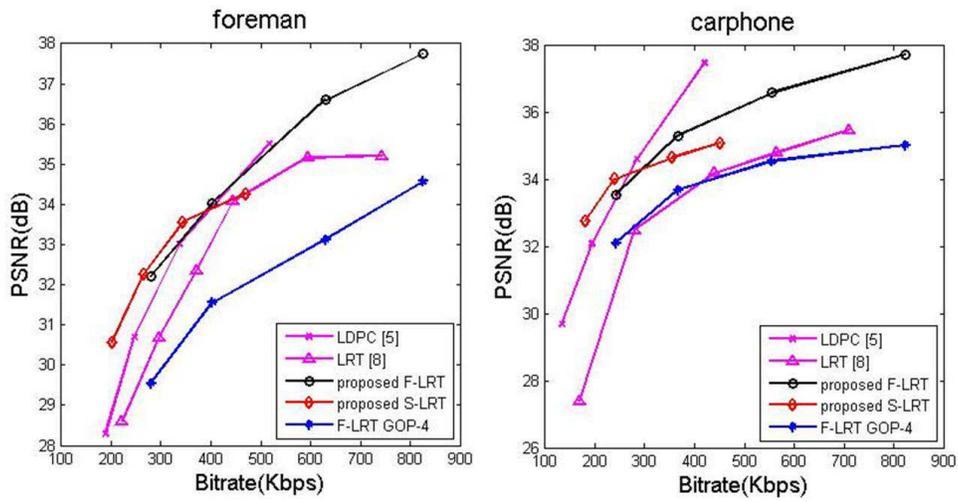}}
 \caption{Average bit rate versus average PSNR comparison of foreman and carphone sequences at 15 frames per second}
 \label{fig:bitrate}
 \end{center}
 \end{figure*}

\section{Results}
Experiments are performed with foreman and carphone QCIF sequences for the first 100 frames at 15 frames per second. For rate adaptation, we used \textit{full LRT} (for high bit rates) and \textit{sampled LRT} (for low bit rates) with N=1,2,3 and 4. After several experiments on different sequences, we noticed that the codec is giving good results at $\delta=-10$ and \textit{step}=2. The bit rate versus PSNR comparison of WZ frames for different schemes are shown in Fig. 9 (15 frames per second). For low bit rates proposed codec performs slightly better compared to LDPC codec. Post processing improves the PSNR drastically for higher neighborhood size. Typical examples of reconstructed WZ frame of different sequences is shown in Fig. 10.  As the DLRTex preserves edges of the image, the proposed codec gives perceptually good reconstructed image.

\begin{figure}[h]
 \begin{center}
 \centering{\includegraphics[width=8cm,height=10.5cm]{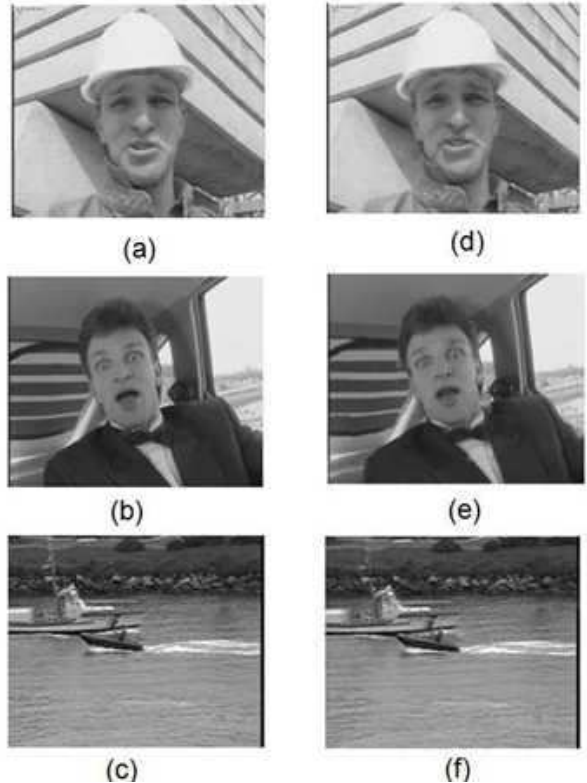}}
 \caption{Reconstructed WZ frame with parameters $N$=2, $\delta=-10$, and $step$=2. a)-c). Original frames, d)-f). reconstructed frames.}
 \label{fig:scheme1}
 \end{center}
 \end{figure}

%\begin{figure*}[hpt]
% \begin{center}
% \centering{\includegraphics[width=16cm,height=8.75cm]{percom2.jpg}}
% \caption{Visual comparison of reconstructed images with original images.($1^{st}$ column- original, $2^{nd}$ column- LRT reconstructed, and $3^{rd}$ column- LDPC reconstructed)}
% \label{fig:scheme1}
% \end{center}
% \end{figure*}

\section{Conclusion}
In this paper we propose a DVC codec based on LRT. At first, we develop an LRT variant to reduce bit rate, namely $\Delta_{odd}$ or $\Delta_{even}$. We also introduce mean assisted ME and MC which contribute towards the improvement of the SI quality. For low bit rates we adopt a strategy, where known and unknown rank pixels are treated differently to improve quality of WZ frame. The comparison of bit rate versus PSNR graphs for different sequences with conventional methods shows that the proposed codec performs near to LDPC schemes and consumes less power. In sampled LRT case, proper up sampling of LRT at the decoder side can improve the PSNR.  As the computational complexity of LRT is much less than LDPC, the proposed codec is suitable for low power applications.


\begin{thebibliography}{1}

\bibitem{1}
J. D. Slepian and J. K. Wolf, ``Noiseless coding of correlated information sources," \emph{IEEE Transactions on
Information Theory}, vol. 19, pp. 471-480, Jul. 1973.

\bibitem{2} A. D. Wyner and J. Ziv, ``The rate-distortion function for source coding with side information at the decoder,"\emph{
IEEE Transactions on Information Theory}, vol. 22, no. 1, pp. 1-10, Jan. 1976.

\bibitem{3}
X. Fan, O. Au, N. Cheung, Y. Chen, and J. Zhou, ``Successive
refinement based Wyner-Ziv video compression,” \emph{Signal
Processing: Image Communication}, vol. 25, pp. 4763, Jan. 2010.

\bibitem{4}
B. Girod,A. Aaron,S. Rane and D. Monedero,`` Distributed video
coding," \emph{Spec Issue Adv Video Coding Deliv}, pp. 71–83, Jan.
2005.

\bibitem{5}
D. Chen et al. ``Unsupervised learning of motion for distributed video coding,'' Online: http://msw3.stanford.edu/~dchen/DVC/
LDPC-Video-DCT-VS-2005.zip, Nov. 2014.

\bibitem{6}
A. M. Aaron, S. Rane, and B. Girod, ``Transform domain Wyner-Ziv codec for video,'' \emph{Proc. SPIE Visual Commun. Image Process,
, Santa Clara, CA}, pp. 520–52, Jan. 2004.

\bibitem{7}
R. Puri, A. Majumdar, and K. Ramchandram, ``PRISM: A video coding paradigm with motion estimation at the decoder," \emph{IEEE Transactions on Image Processing}, vol. 16, no. 10, pp. 2436-2448, Oct. 2007.

\bibitem{8}
K. Mallick and J. Mukherjee, ``Distributed Video Coding using Local Rank Transform,'' \emph{Indian Conference on Computer Vision, Graphics and Image Processing}, Dec. 2014.

\bibitem{9}
J. Mukherjee, ``Local rank transform: Properties and applications,'' \emph{Pattern Recognition Letters}, \emph{Elsevier}, vol. 32, pp. 1001-1008, 2011.

\bibitem{10}
M. Ahmadvand and A. Ezhdehakosh,``A new pipelined architecture for JPEG2000 MQ-coder,'' \emph{proc. World Cong on Engg and Compo Sci.}, vol. 2, Oct. 2008.

\bibitem{11}
M. Dyer, S. Nooshabadi, D. Taubman, ``Design and analysis of system on a chip encoder for JPEG 2000,'' \emph{IEEE Transactions on Circuits and Systems for Video Technology}, vol. 19(2), pp.215-225, 2009.

\bibitem{12}
A. Fog, Instruction tables lists of instruction latencies, throughputs and microoperation breakdowns for Intel, AMD and VIA CPUs, Online$:http://www.agner.org/optimize/instruction_tables.pdf$, Apr. 2013, Technical University of Denmark.

\end{thebibliography}
\end{document}